\documentclass[aps,prb,twocolumn,nofootinbib,superscriptaddress,showpacs,preprintnumbers,notitlepage,floatfix]{revtex4-2}

\usepackage{amsmath,amsfonts,amssymb}
\usepackage[utf8]{inputenc}
\usepackage{braket}
\usepackage{graphicx}
\usepackage{bm}
\usepackage{placeins}

\usepackage{xfrac}
\usepackage[capitalize]{cleveref}
\usepackage{changes}
 
\usepackage{ bbold }
\usepackage{xspace}
\let\revappendix\appendix

\newcommand{\h}[1]{\hat{#1}}
\newcommand{\wh}[1]{\widehat{#1}}
\newcommand{\spin}{s} 
\renewcommand{\vec}[1]{\boldsymbol{#1}}

\newcommand{\jj}{$J_1$-$J_2$\xspace}

\newcommand{\Jmax}{\ensuremath{j_{\textrm{max}}}}

\newcommand{\linktomain}[1]{#1}

\begin{document}
\title{Many-Body Quantum States with Exact Conservation of Non-Abelian and Lattice Symmetries through Variational Monte Carlo}

\author{Tom Vieijra}
\thanks{These two authors contributed equally}
\affiliation{Department of Physics and Astronomy, Ghent University, 9000 Ghent, Belgium}
\author{Jannes Nys}
\thanks{These two authors contributed equally}
\affiliation{University of Antwerp - imec, IDLab - Faculty of Applied Engineering Sint-Pietersvliet 7, 2000 Antwerp, Belgium}
\affiliation{Department of Physics and Astronomy, Ghent University, 9000 Ghent, Belgium}

\begin{abstract}
Optimization of quantum states using the variational principle has recently seen an upsurge due to developments of increasingly expressive wave functions. In order to improve on the accuracy of the ans\"atze, it is a time-honored strategy to impose the systems' symmetries. We present an ansatz where global non-abelian symmetries are inherently embedded in its structure. We extend the model to incorporate lattice symmetries as well. We consider the prototypical example of the frustrated two-dimensional \jj model on a square lattice, for which eigenstates have been hard to model variationally. Our novel approach guarantees that the obtained ground state will have total spin zero. Benchmarks on the 2D \jj model demonstrate its state-of-the-art performance in representing the ground state. Furthermore, our methodology permits to find the wave functions of excited states with definite quantum numbers associated to the considered symmetries (including the non-abelian ones), without modifying the architecture of the network.
\end{abstract}

\maketitle

\section{Introduction} 
Variational optimization of quantum many-body states to capture the low-lying eigenstates of quantum many-body systems is increasingly being studied as a way to uncover properties and mechanisms of physical systems. Many algorithms and variational wave functions have been put forward as viable candidates to describe many-body systems efficiently.  Well-known examples include the use of Slater determinants for fermionic systems \cite{slater1929the}, the density matrix renormalization group (DMRG) and its extensions to multiple dimensions \cite{schollwock2011the, orus2019tensor}, the resonating valence bond wave functions for frustration and superconductivity \cite{anderson1987the} and many more.  Recently, a new class of wave functions has emerged, which are based on the success of neural networks in machine learning and are optimized by sampling basis states of the quantum mechanical problem~\cite{Carleo17}.

Many physical systems have underlying symmetries encoded in their Hamiltonian.  
Respecting the symmetries of the system is important for various applications, including quantum chemistry~\cite{szabo2012modern, gunst2019three} and the description of spin liquids~\cite{balents2010spin, savary2016quantum}.
Both lattice symmetries and on-site symmetries yield important constraints on the wave function, and can be harnessed to gain insight in the wave function and to gain efficiency in its optimization.  Having control over how the variational wave function transforms under symmetry operations gives access to associated quantum numbers, which are useful to characterize the spectrum of many-body Hamiltonians. Constructing wave functions that transform in a well-defined way under symmetry operations is a non-trivial task, but has led to very efficient and accurate results in the past. In the class of tensor networks, symmetries have been encoded in their tensorial structure, leading to more efficient algorithms and insight into the physical system \cite{Singh2010tensor, Singh2012tensor, schmoll2020programming, orus2019tensor}.
In the case of neural network wave functions, advances have been made to include abelian symmetries~\cite{Choo18} and more recently non-abelian continuous spin symmetries in one dimension~\cite{vieijra2020restricted}. More generally, the question of how to include symmetries in deep neural networks is an active field of research, which has led to neural network architectures which transform equivariantly under symmetry operations of the input~\cite{cohen2016group}. 

In this paper, we combine insight from equivariant neural networks and tensor networks to construct a variational wave function that transforms well under \emph{both} global continuous non-abelian symmetries and discrete lattice symmetries. We benchmark our construction numerically on the two-dimensional \jj model, which has proven difficult to model using conventional neural networks due to breaking of the non-abelian SU(2) symmetry \cite{Choo18, ferrari2019neural}. 

\section{Approach} 
To find eigenstates of quantum many-body systems, we introduce a variational wave function built entirely of equivariant operations with respect to non-abelian SU(2) symmetry, as well as a set of lattice symmetries.  Equivariance of an operation $f(x)$ under a symmetry group $G$ is defined as $f(x)$ having the property 
\begin{equation}
    g \cdot f(x) = f(g \cdot x) \quad \forall g \in G.
    \label{eq:equivariance}
\end{equation} 
The construction of our SU(2) equivariant network is inspired by recent progress in deep learning~\cite{kondor2018clebsch, cohen2016group}, where group equivariant neural networks were introduced.  In the same vein as Ref.~\cite{cohen2016group}, the equivariance under lattice symmetries can be trivially fulfilled, as we will show below.
For the SU(2) symmetry, however, an additional difficulty compared to Ref.~\cite{kondor2018clebsch} is that we need to obtain representations of the whole many-body state with definite \emph{total} angular-momentum, which we accomplish by defining a tree-like coupling scheme to perform Clebsch-Gordan transformations. This bears resemblance to tree tensor networks, where one uses a network of tensors where iteratively two (composite) systems are coupled into a single composite system with definite angular momentum in a tree-like fashion~\cite{gunst2019three}.
However, our network is trained via Variational Monte Carlo, where the parameter gradients are computed through automatic differentiation. The Monte Carlo approach to variational wave functions also allows us to make the ansatz have well-defined quantum numbers with respect to lattice symmetries by defining symmetrized probability amplitudes \cite{nomura2020helping}.

To outline our network construction, we consider $N$ spins on a lattice, where every spin can be labeled with an angular momentum quantum number $s_i$.  
The individual spin states are vectors in the single particle Hilbert space $\mathcal{H}_i$, which can be described in a basis of size $2s_i+1$. For example, one can choose the basis of spin projections $m_{s_i}$ along a quantization axis.  
The direct product of these one-particle Hilbert spaces $\mathcal{H} = \mathcal{H}_1 \otimes .... \otimes \mathcal{H}_N$ forms the Hilbert space of the composite system.  
A state $\ket{\Psi}$ in $\mathcal{H}$ can thus be expanded in a basis consisting of all possible combinations of single-particle basis states
\begin{equation}
    \ket{\Psi} = \sum_{\mathcal{S}}{\Psi(m_{s_1},...,m_{s_N}) \ket{s_1,m_{s_1};...;s_N,m_{s_N}}},
    \label{eq:expansion}
\end{equation}
where $\Psi(m_{s_1},...,m_{s_N})$ is the expansion coefficient of $\ket{\Psi}$ in the basis $\ket{s_1,m_{s_1};...;s_N,m_{s_N}}$. For brevity, we will use the notation $\mathcal{S} = (m_{s_1},...,m_{s_N})$ to denote a basis state.
The eigenstates $\ket{\Psi}$ inherit the symmetries from the related  Hamiltonian $\wh{H}$ of the system.
Generally, the Hamiltonian commutes with operators of symmetry groups acting on the degrees of freedom $\widehat{U}(g)$, with $g \in G$, where $G$ is a symmetry group.  
In those cases, the states $\ket{\Psi}$ should also be eigenstates of the symmetry operators.

\subsection{SU(2) symmetry}
For SU(2), the action of symmetry operations can be written as $\wh{U}(\bm{\theta}) = e^{i(\theta_x \h{s}_x + \theta_y \h{s}_y + \theta_z \h{s}_z)} = e^{i\bm{\theta} \cdot \h{\mathbf{s}}}$, where $\bm{\theta}$ is a vector in three dimensional space denoting a rotation of magnitude $\theta$ around the axis defined as the direction of $\bm{\theta}$ and $\h{\mathbf{s}}$ is a vector of the generators of SU(2).  For spin-$\frac{1}{2}$ degrees of freedom, these are the Pauli matrices.  For a composite system, the symmetry action is the direct product of the single-particle actions $\wh{U}(\bm{\theta}) = e^{i\bm{\theta} \cdot (\h{\mathbf{s}}_1 + ... + \h{\mathbf{s}}_N)} = e^{i\bm{\theta} \cdot \h{\mathbf{S}}}$, with $\h{\mathbf{S}}$ the total angular momentum operator.  To make sure that the wave function of a system of $N$ particles is an eigenstate of the symmetry action, it should have well-defined angular momentum.  

We will form the expansion coefficients $\Psi(m_{s_1},...,m_{s_N})$ through a sequence of operations (from hereon called layers), in order to obtain a variational wave function with a well-defined total angular momentum.
In every layer we construct composite systems that transform as an irreducible representation with respect to SU(2), meaning that they can be labeled by an angular momentum quantum number. In this way, we gradually build a state that encompasses the whole system of $N$ degrees of freedom, while keeping a structure of definite angular momentum.

\begin{figure*}[tb]
    \centering
    \includegraphics[width=0.9\textwidth]{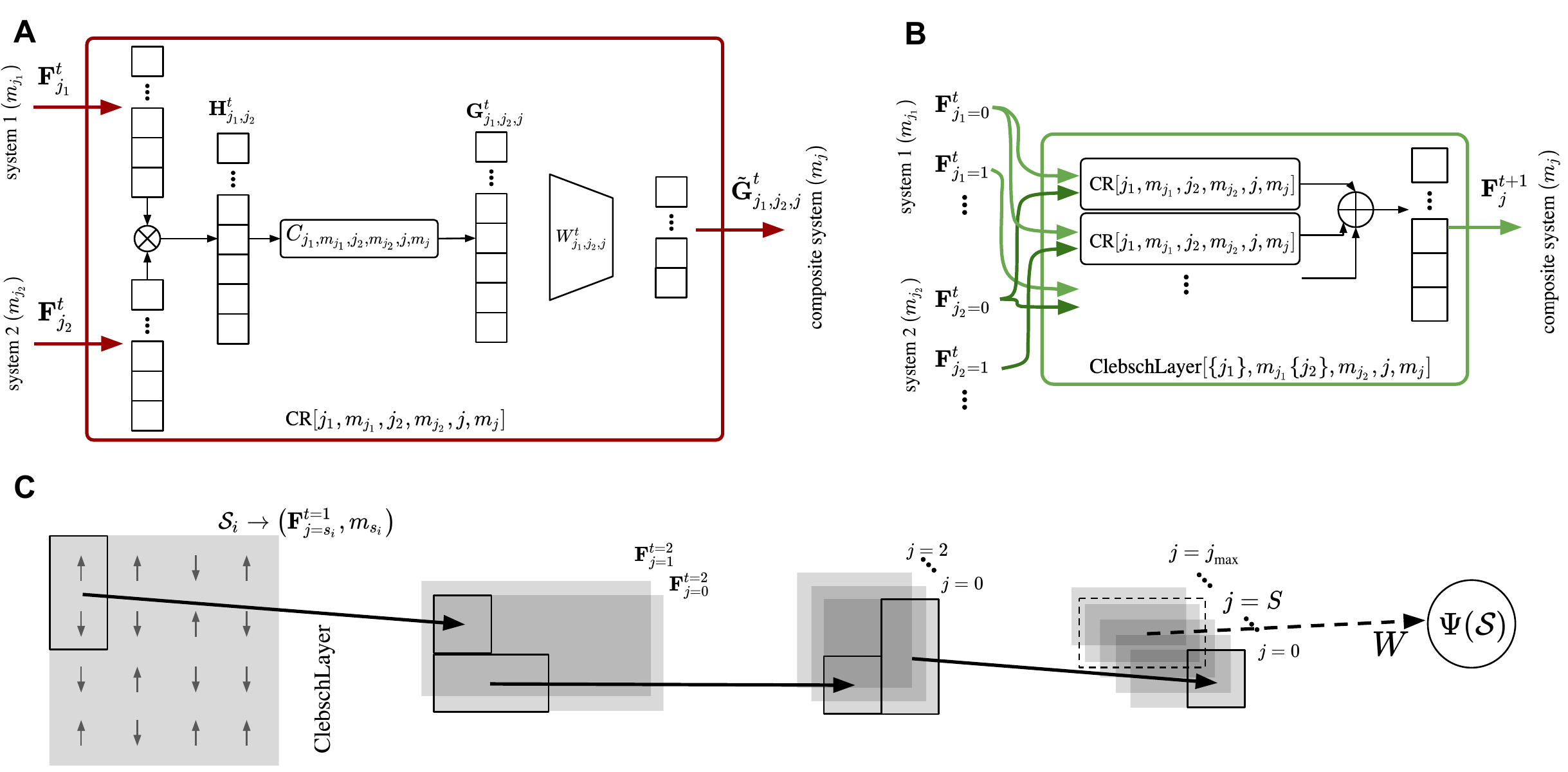}
    \caption{Illustration of the architecture. The bottom panel (\textbf{C}) illustrates the flow of the computation. The network receives a spin configuration $\mathcal{S}$, with components $(\spin_i, \mathcal{S}_i = m_{s_i})$, from which one can infer $(\mathbf{F}^{t=1}_{j=\spin_i}, m_{s_i})$ at each site $i$. As indicated by the rectangle, the ClebschLayer takes as an input two vectors $\mathbf{F}^{t=1}_{\spin_i}$ and $\mathbf{F}^{t=1}_{\spin_j}$ from different sites (here, along the vertical direction), and outputs two new definite angular-momentum vectors $\mathbf{F}_{j}^{t=2}$ (with $j=0,1$) to the next layer. Similarly, in the subsequent layers $t$, the vectors $\mathbf{F}^t_{j_1}$ and $\mathbf{F}^t_{j_2}$ (from different composite systems) are coupled to obtain the new definite angular-momentum vector $\mathbf{F}^{t+1}_{j}$. At layer $D = \log_2(N)$, we only compute $F_{S}^{D}$ with the desired total angular momentum $S$. After a last dimensionality reduction indicated by $W$, we obtain $\Psi(\mathcal{S})$.
    (top left, (\textbf{A})) The ClebschReduction (CR) module takes as an input vectors from neighboring sites. Sequentially, we compute tensor $\mathbf{H}$ from their tensor product ($\otimes$), which is then projected onto definite $j$ via the Clebsch-Gordan coefficient $C_{j_1,m_{j_1},j_2,m_{j_2},j,m_{j}}$ to obtain a vector $\mathbf{G}$ with a hidden dimension $\tau_{j_1}^t\tau_{j_2}^t$. The latter dimension is reduced to $\tau_{j}^t$ via a parameter matrix $W_{j_1, j_2, j}^t$. (top right, (\textbf{B})) The ClebschLayer combines multiple CR modules to compute the $\mathbf{F}_{j}^{t+1}$, for all input angular momenta that can be coupled to obtain $j$ (i.e.\ set $\mathcal{J}$ in the main text). The vectors $\mathbf{G}$, stemming from the different input angular momenta, are added elementwise ($\oplus$) to obtain the final $\mathbf{F}_{j}^{t+1}$. 
    }
    \label{fig:fig_coupling}
\end{figure*}

In the following, we describe the action of a layer in our variational wave function.  To make the following description clearer, we assume a square lattice in two dimensions, but the construction can be generalized to other dimensions and other topologies.  We first define an operation which performs pairwise coupling of angular momenta (see Fig.~\ref{fig:fig_coupling}A).  The pattern in which pairs are taken can be chosen freely, for example, one could pair horizontally or vertically neighboring angular momenta, alternating the direction at every layer (see Fig.~\ref{fig:fig_coupling}C).  

The input variables of a layer are defined on a lattice, where we define vectors $\mathbf{F}^{t}_{j} \in \mathbb{R}^{\tau_j^t}$ at every point, associated to each possible angular momentum $j$, and $t$ denotes the considered layer (where $t=1$ is the initial spin configuration). 
For a given configuration $\mathcal{S}$,  $\mathbf{F}^{t}_{j}$ represents a (variational) weight vector of a composite subsystem of $2^{t-1}$ spins that are coupled to a definite total angular momentum $j$, as shown in Fig.~\ref{fig:fig_coupling} C. The corresponding $m_j$ is the sum of the individual spin projection quantum numbers in the composite subsystem. The dimension $\tau_j^t$ is a measure that determines the variational freedom that is used to describe a composite subsystem (at layer $t$) with angular momentum $j$.  In the most general case, $\tau_{j}^t$ may depend on the angular momentum $j$, the lattice position and the layer index $t$. We will use $\tau_j$ and $\tau_j^t$ interchangeably below. We do not include a spatial dependence in our numerical calculations due to lattice symmetry considerations.  The physical lattice of $N$ spin-degrees of freedom $s$ is represented in our notation as $\mathbf{F}_{j=\spin}^{t=1} = [1]$  (where $\tau_{j=\spin}^{t=1}=1$) for each of the $N$ sites. For a given configuration $\mathcal{S}$, we can assign a spin-projection quantum number to each site. 

Having defined the structure of the input, the next step is to apply Clebsch-Gordan transformations to the angular momenta of neighboring sites according to the scheme detailed in Fig.~\ref{fig:fig_coupling}. 
% This will yield new vectors, $\mathbf{F}^{t+1}_{j}$.
The coupling of angular momenta into composite subsystems is visualized in Fig.\ref{fig:fig_coupling}A. We take the tensor product of two $\mathbf{F}$ vectors  corresponding to a pair of angular momenta $j_1$ and $j_2$ at neighboring positions, and obtain a vector $\mathbf{H}^t_{j_1, j_2} = \mathbf{F}^t_{j_1} \otimes \mathbf{F}^t_{j_2}$ with dimension $\tau_{j_1}^t\tau_{j_2}^t$.  Since $\mathbf{H}$ does not have well-defined angular momentum (e.g.~$\frac{1}{2}\otimes \frac{1}{2}\to 1\oplus 0$ results in a mixture of angular momenta $j=0$ and $j=1$), it is projected onto quantum number $j$ via the Clebsch-Gordan coefficients $C$ to obtain a new vector $\mathbf{G}$ as follows
\begin{equation}
    \mathbf{G}^t_{j_1, j_2, j} = C_{j_1, m_{j_1}, j_2, m_{j_2},j,m_j} \times \mathbf{H}^t_{j_1, j_2},
\end{equation}
where $m_{j_1}$ and $m_{j_2}$ are the angular momentum projections of the original composite subsystems. These composite subsystems (or regions of the original spin lattice) should be non-overlapping and $m_j$ is uniquely obtained as $m_j = m_{j_1} + m_{j_2}$.  In this way, we constructed a new lattice with half the sites of the lattice of input variables (see Fig.~\ref{fig:fig_coupling}C).

To control the dimension $\tau_{j_1}^t \tau_{j_2}^t$ of the vector $\mathbf{G}^t_{j_1, j_2, j}$, we apply a linear transformation mapping the $\tau_{j_1}^t \tau_{j_2}^t$ dimensions to $\tau_j^{t+1}$ linear combinations of them.  This linear transformation is performed by a parameter matrix $W_{j_1, j_2, j}^{t}$ and introduces a dimensionality reduction. We point out that $W_{j_1, j_2, j}^{t}$ does not depend on the angular-momentum projections $m_{j_1}$, $m_{j_2}$ and $m_{j}$, in accordance with the Wigner-Eckart theorem.
We obtain the vector
\begin{align}\label{eq:G}
\mathbf{\tilde{G}}^t_{j_1, j_2, j} = W_{j_1, j_2, j}^{t} \times \mathbf{G}^t_{j_1, j_2, j},  
\end{align}
where $W_{j_1, j_2, j}^{t}$ has $\tau_j^{t+1}$ rows, while the number of columns is $\tau^t_{j_1} \tau^t_{j_2}$. 
This procedure of obtaining a vector $\mathbf{\tilde{G}}$ consisting of states with well-defined angular momentum $j$ is repeated at layer $t$ for all $(j_1,j_2)$ from the set $\mathcal{J} = \{(j_1, j_2) \mid |j_1-j_2| \leq j \leq j_1+j_2\}$.  We now combine all vectors $\mathbf{\tilde{G}}$ with angular momentum $j$ by summing them element wise in the hidden dimension into a single vector depending only on the final angular momentum $j$ (see Fig.~\ref{fig:fig_coupling}B).  This procedure of summing the vectors can also be seen as finding $\tau_j^{t+1}$ linear combinations of the $\bar{\tau}_{j}^t = \sum_{(j_1, j_2) \in \mathcal{J}} \tau_{j_1}^t \tau_{j_2}^t$ features with angular momentum $j$, resulting from all possible combinations of $(j_1,j_2)$.
This results in the vector $\mathbf{F}_{j}^{t+1}$, ready to be used as input for the next layer of our network.
This procedure (called `ClebschLayer' in Fig.~\ref{fig:fig_coupling}B) provides us with a building block to construct variational wave functions with well-defined total angular momentum, with variational parameters given by all weight matrices $W$. We take $\tau_j^{t+1}$ (which determines dimensions of $W_{j_1,j_2,j}^t$) to be the minimum of $\bar{\tau}_{j}^{t}$ and a predefined $\tau^{t+1, \textrm{max}}_j$. By repeating the layer described above $\log_2(N)$ times (see Fig.~\ref{fig:fig_coupling}),
we essentially couple the individual angular momenta to a well-defined total angular momentum $S$.  After carrying out a final dimensionality reduction, we obtain an estimate for $\Psi(\mathcal{S})$.

\subsection{Lattice symmetries}
Having defined a wave function which, given a sample  of configurations of spin projections, yields a probability amplitude $\Psi_{S,M_S}(\mathcal{S})$ with well-defined angular momentum $S$ and angular momentum projection $M_S$, we can now improve the ansatz by including discrete lattice symmetries such as rotations, reflections or translations.
Restricting the considered Hilbert space (for non-SU(2) conserving ans\"atze) to eigenstates of the lattice symmetries has been shown to improve the obtained ground state energies by multiple orders of magnitude~\cite{nomura2020helping}.  Instead of defining the probability amplitude as $\Psi_{S,M_S}(\mathcal{S})$, we define it as a linear combination of amplitudes with configurations transformed according to the (discrete) symmetry group $G$~\cite{Choo18,seki2020symmetry}:
\begin{equation}\label{eq:lattice_symmetries}
    \Psi_{S,M_S}^{\kappa}(\mathcal{S}) = \frac{1}{\left|G\right|}\sum_{g \in G} \kappa_g \Psi_{S,M_S}(g(\mathcal{S})),
\end{equation}
where $\kappa_g$ is the character associated to group element $g$ and irreducible representation $\kappa$. In practice, we compute every elementary internal operation in our method together with its operation on all input configurations transformed by $g \in G$ 
(weighted by $\kappa_g$). In this way, every operation is made trivially equivariant under the lattice symmetries. Unless specified otherwise, we consider the trivial irreps of the symmetry groups, i.e. the irrep with momentum $\mathbf{k} = (0,0)$ of the group of translations and the irrep $A_1$ of the $C_{4v}$ group of rotations and reflections.

\medskip
Through this approach, we can optimize our network using Variational Monte Carlo (VMC) and Stochastic Reconfiguration \cite{Sorella98}. Hereby, we use Markov Chain Monte Carlo to generate $N_{samples}$ configurations $\mathcal{S}$ according to the probability distribution $\left| \Psi(\mathcal{S}) \right|^2$.  With these samples, estimates of the energy and gradients with respect to the variational parameters can be obtained.  The variational parameters are then updated by first solving the linear equations $\mathbf{S} \mathbf{x} = \mathbf{g}$ for $\mathbf{x}$, where $\mathbf{S}$ is the correlation matrix of the gradient of $\log{\Psi(\mathcal{S})}$ and $\mathbf{g}$ is the vector of the gradients of the energy with respect to the variational parameters~\cite{Carleo17, carleo2019netket}.  When $\mathbf{x}$ is found, we use the Adam optimizer with learning rate $\eta$
%and momentum rates $\beta_1=0.9$ and $\beta_2=0.999$} 
to determine an update of the variational parameters~\cite{kingma2014adam}. We iterate this procedure until convergence is reached. For details of the optimization procedure, see Appendix~\ref{app:optimization}.

\subsection{Relation to other work}
In previous work on conserving SU(2)-symmetry through variational ans\"atze in the context of VMC \cite{vieijra2020restricted}, one works with configurations in the space of irreducible representations of subsystems.  In particular, considering a 1D chain for clarity, one chooses the basis set defined by the set of all possible combinations $\{j_1,...,j_N\}$, where $j_i$ is the total angular momentum of the first $i$ spins. 
One could view this ansatz as applying a mapping from the space of angular momenta to its probability amplitudes via a variational wave function.  The approach we outlined here adopts a completely different viewpoint, where the mapping to sets of angular momenta $\{j_1,...,j_N\}$ is done in the variational wavefunction \emph{itself}.  Adopting this viewpoint, our network considers spin configurations $\mathcal{S}$, and maps this to its probability amplitude by building composite configurations $\{j_1,...,j_N\}$ with well-defined angular momenta in the ansatz, and deciding on their importance using variational parameters.  This construction allows to efficiently calculate local observables, since we still have access to the spin configurations, in contrast to the approach from Ref.~\cite{vieijra2020restricted}, where one has to define local observables in the space of angular momenta. Doing this can become problematic in dimensions larger than one, where operators in the space of angular momenta become increasingly non-local.  Another advantage is that access to the spin configurations allows to efficiently include lattice symmetries as shown in Eq.~\eqref{eq:lattice_symmetries}, as opposed to the previous work. Indeed, in the previous approach, the action of an element of the lattice symmetries does not map the $\{j_1, ..., j_N\}$ basis states to another one, but rather to a non-trivial linear combination of them. As we will discuss in the next sections, including lattice symmetries can increase the accuracy of our wave functions by orders of magnitude.

\section{\jj model} 
The \jj model is a prototypical model with frustrated interactions. The Hamiltonian
\begin{align}
    \wh{H} = J_1 \sum_{\langle ij \rangle} \h{\vec{s}}_i . \h{\vec{s}}_j + J_2 \sum_{\langle\!\langle ij \rangle\!\rangle} \h{\vec{s}}_i . \h{\vec{s}}_j
\end{align}
contains a nearest neighbor ($J_1$) and next-nearest-neighbor ($J_2$) Heisenberg interaction term. Here, $\h{\vec{s}}_i$ denotes the spin operator at site $i$, and $\langle ij \rangle$ ($\langle\!\langle ij \rangle\!\rangle$) denotes the set of all nearest (next-nearest) pairs.
Note that both terms conserve SU(2) symmetry, and therefore, the eigenstates are SU(2) symmetric as well.
In the limit of $J_2 \rightarrow 0$, the model exhibits a N\'eel-type antiferromagnetic order, while in the limit of $J_1 \rightarrow 0$, a stripe-type antiferromagnetic order is found. In the intermediate regime where $J_2/J_1 \approx 0.5$, the system is highly frustrated, which makes this system a hard problem to solve by contemporary methods, and is thus widely used as a benchmark for novel algorithms~\cite{choo2019two,nomura2020dirac}. 
There are numerous conflicting proposals with regard to the structure of the ground state, including the plaquette valence-bond state~\cite{zhitomirsky1996valence, yu2012spin}, the columnar valence-bond state~\cite{sachdev1990bond, singh1999dimer} or a gapless spin liquid~\cite{capriotti2001resonating, liu2018gapless}, but a conclusive characterization of the phase diagram is still missing.

\section{Results} 

\begin{figure}[tb]
    \centering
    \includegraphics[width=0.5\textwidth]{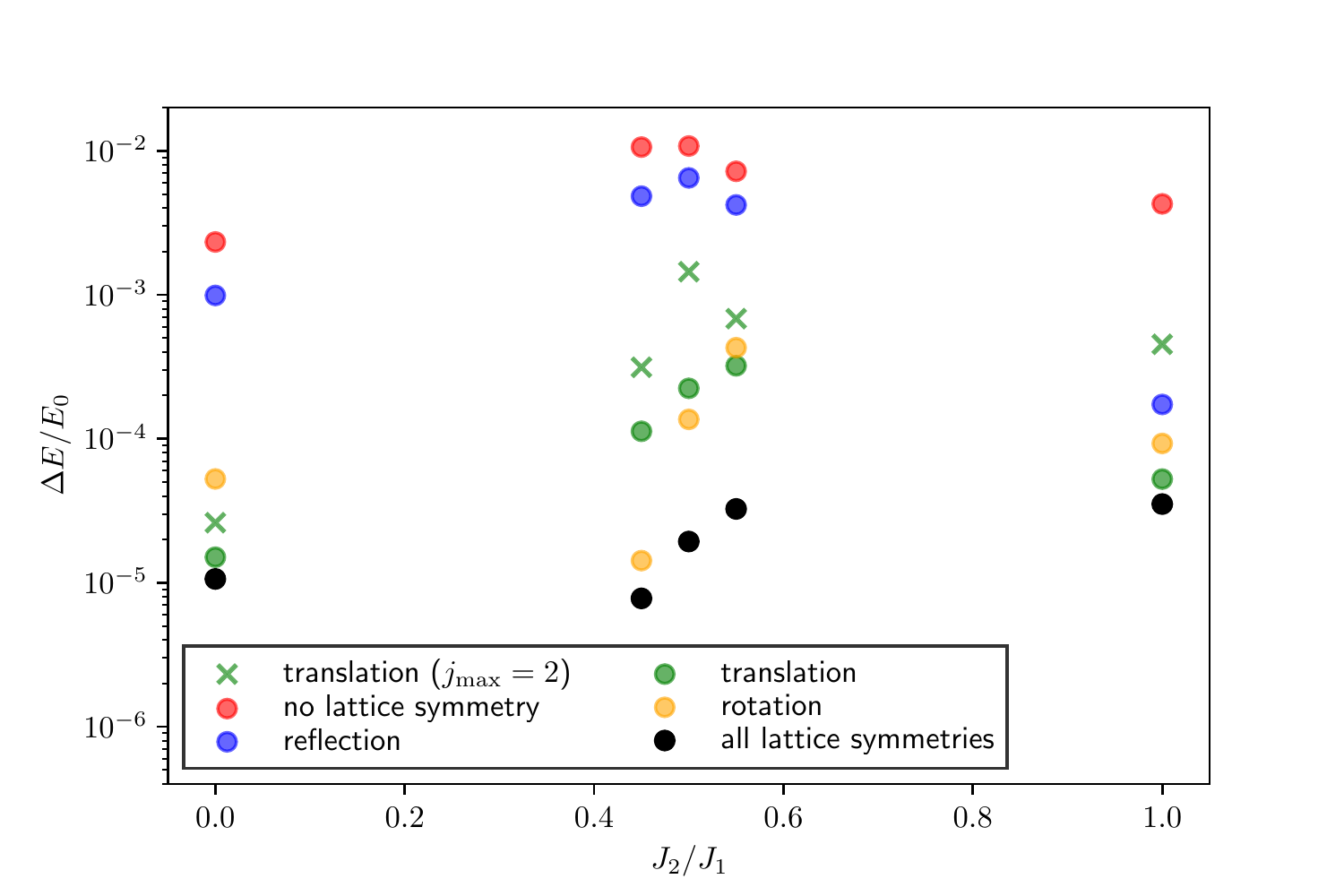}
    \caption{Relative error on the ground state energy $\Delta E/E_0 = (E_{0,\textrm{exact}} - E)/E_{0,\textrm{exact}}$ of the \jj model on a $4\times 4$ lattice. We illustrate the results for a model with rotation, translation, or reflection symmetry imposed through Eq.~\eqref{eq:lattice_symmetries} for $\Jmax=4$. We also show the effect of $\Jmax=2$ for the translation symmetric model. Also shown are the results for a model with all symmetries and only SU(2) symmetry. The exact ground state energy $E_{0,\textrm{exact}}$ is obtained through exact diagonalization.\label{fig:4x4_symmetries}}
\end{figure}

\subsection{Effect of lattice symmetries}
The ansatz proposed in this work conserves SU(2) through its architecture, while lattice symmetries are imposed through Eq.~\eqref{eq:lattice_symmetries}. Therefore, we start by investigating the effect of the latter on the obtained ground state energy.  We use our model with a homogeneous hidden dimension $\tau=8$ for all layers and angular momenta.  We optimize our models through VMC with learning rate $\eta=0.001$ and $N_{samples}=4096$.
Figure~\ref{fig:4x4_symmetries} shows the computed ground-state energy of the \jj model at different ratios $J_2/J_1$ on a $4\times 4$ lattice with periodic boundary conditions (PBC). We observe that lattice symmetries improve the relative error by over two orders of magnitude. Especially the incorporation of translation and rotation symmetry tend to result in the lowest ground-state energies. Near $J_2/J_1=1$, where next-to-nearest-neighbor interactions dominate, however, the impact of translation symmetry is most prominent. We also point out that the effect of the intermediary angular-momentum cutoff $\Jmax$ is most prominent near this point, which can be observed by comparing the $\Jmax=2$ and $\Jmax=4$ results in Fig.~\ref{fig:4x4_symmetries}. The latter is expected, since the singlets become more long ranged, and therefore require the propagation of more angular momentum information higher up the tree structure.

\subsection{Systematic effects of hyper-parameters}\label{sec:hyperparams}
To show how our method is systematically improvable, we investigate how the energy improves with increasing number of variational parameters. Specifically, we optimize our network for the ground state of the \jj model on an $8\times 8$ lattice, where we take $J_1=1$ and $J_2=0.5$. We optimize networks with all symmetries (i.e. SU(2) complemented with translation, reflection and rotation symmetries) and a Markov Chain Monte Carlo algorithm with nearest neighbour exchange from which we take $N_{samples} = 1024$. We use a learning rate of $\eta=0.005$.  For every run, we let the algorithm update the variational parameters $1000$ times.  Here, we consider the hidden dimension $\tau_j^t$ to be homogeneous for all layers and all angular momenta, and denote it with $\tau$.

\begin{figure}[bth]
    \centering
    \includegraphics[width=0.5\textwidth]{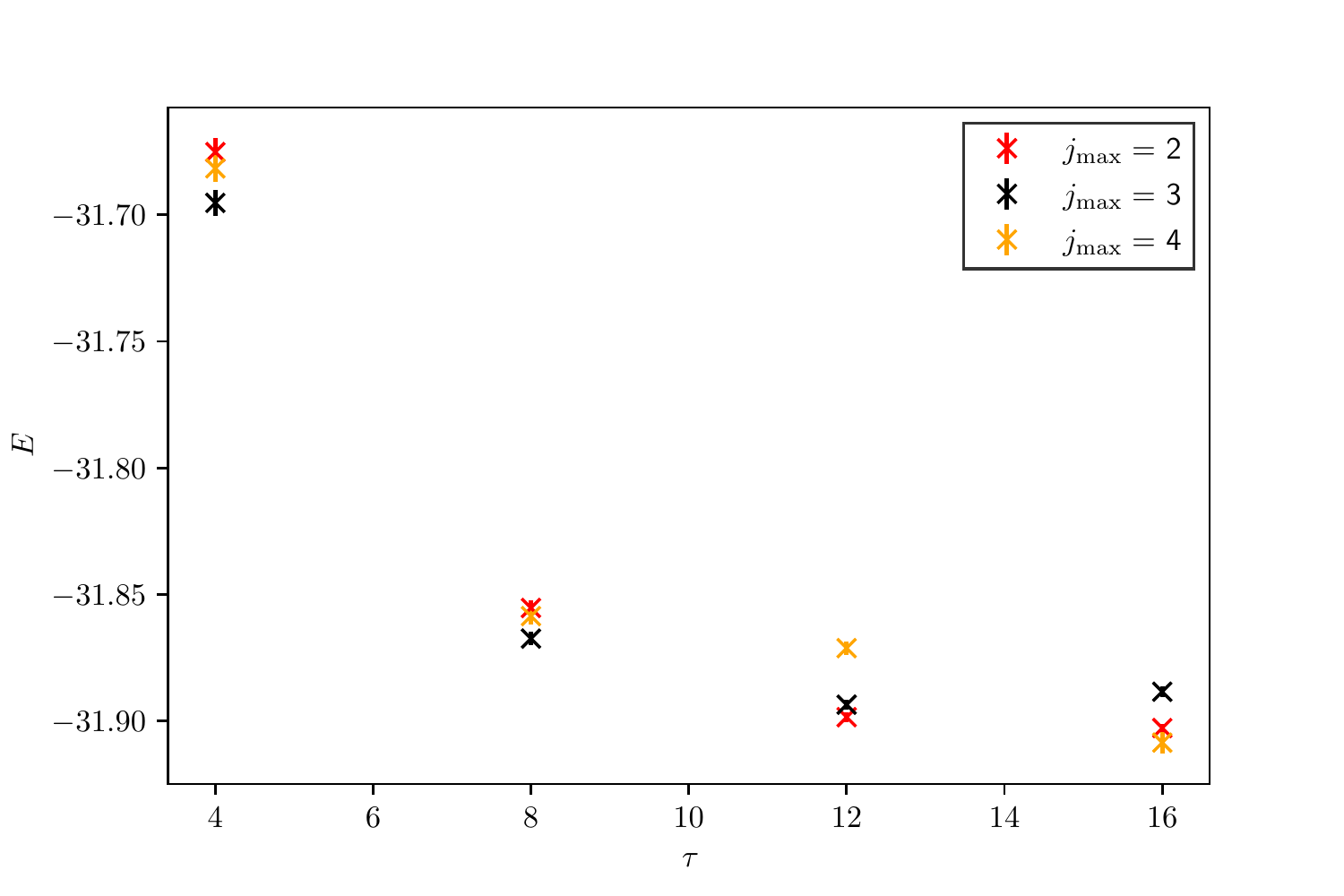}
    \caption{Variation of the obtained ground state energy $E$ of an $8 \times 8$ \jj model with $J_1=1$ and $J_2=0.5$ as a function of the hyperparameters $\Jmax$ (angular-momentum cutoff), and the hidden dimension $\tau$.\label{fig:8x8_E_ifo_tau}}
    \label{fig:tau_dependence}
\end{figure}

The number of variational parameters is determined by the values of the maximum hidden dimension $\tau$ and the cut-off $\Jmax$ of the intermediate angular momenta. Typically, $\Jmax \leq 4$ is sufficient to obtain a good approximation of the systems considered in this work.
We study the effect of these parameters in Fig.~\ref{fig:8x8_E_ifo_tau}, and observe that the value of $\tau$ is most important in achieving low energies. At high $\tau$, the accuracy of the energy depends on the propagation of large momenta between composite systems that are spatially separated.  

\begin{figure}[bt]
    \centering
    \includegraphics[width=0.5\textwidth]{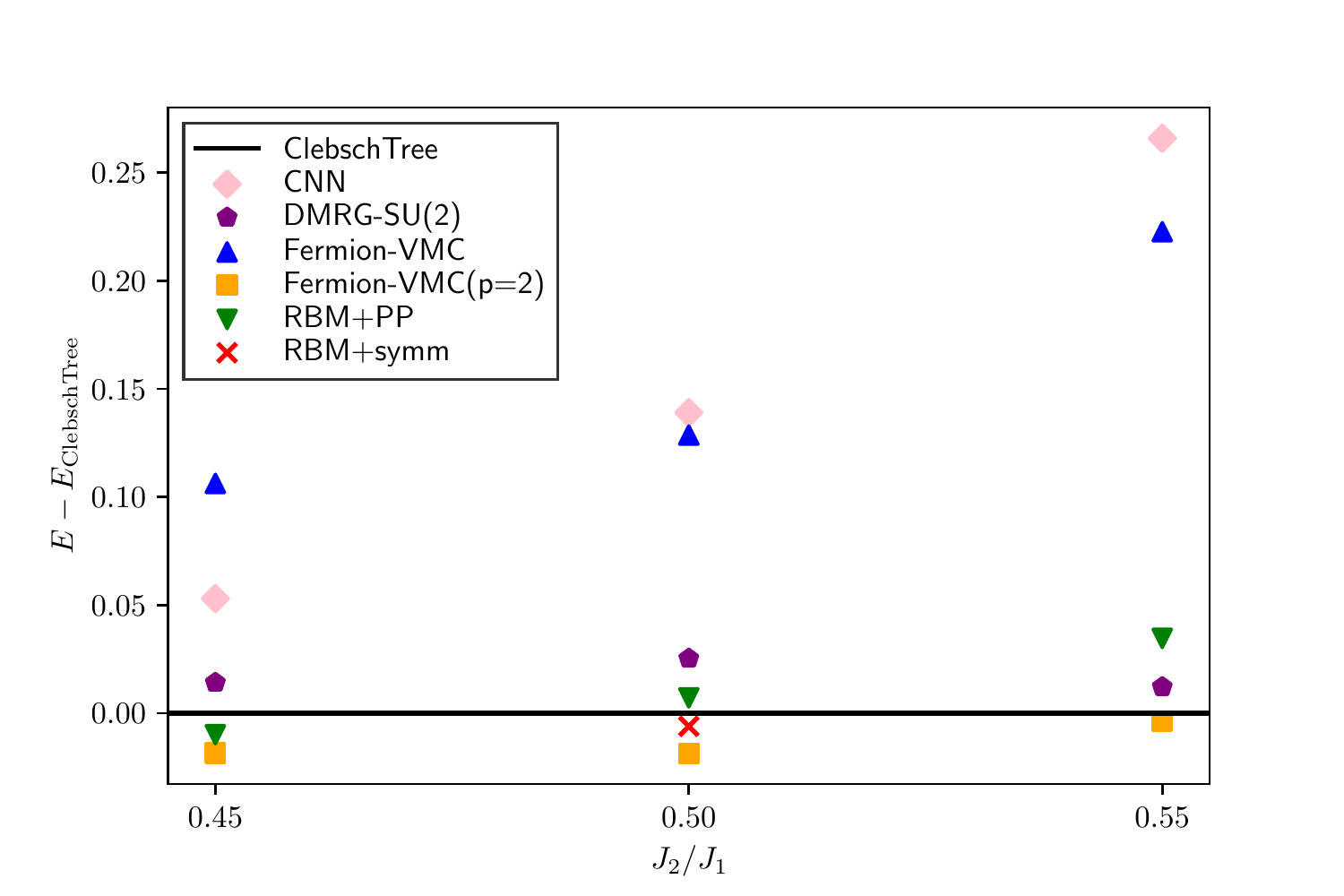}   
    \caption{Ground state energy difference for $8 \times 8$ lattices with various $J_2/J_1$, determined using various ans\"atze. 
    For all methods, we subtract the ground state energy obtained with our model.\label{fig:results8x8}}
\end{figure}

\subsection{Comparison with other approaches}
Knowing that the variational energy saturates at $\tau=16$ and $\Jmax=4$, we refined the variational content of the wave function by introducing a dependence in $\tau_j^t$ on the layer and the angular momentum. The optimal values were obtained through a gridsearch with a low number of samples. For our benchmark simulations on the $8 \times 8$ lattice, we found that $\tau_{j \leq 2}^{t, \textrm{max}} = 16$ and $\tau_{j > 2}^{t, \textrm{max}} = 8$, and $\Jmax=4$ yield good accuracies wile minimizing the number of variational parameters.  %For the remainder of this work, we take these hyperparameters of the model.

We benchmark our method on the $8 \times 8$ lattice with PBC, and compare to other available methods in Fig.~\ref{fig:results8x8}. The numerical values for the energies, details on the optimization procedure and hyperparameters can be found in Appendices~\ref{app:numval} and~\ref{app:optimization}.
To determine the effect of an architecture which respects SU(2) by construction, we compare our results to the CNN ansatz from Ref.~\cite{choo2019two}, which does not conserve SU(2). In the latter, a spike in relative error is observed near $J_2/J_1=0.55$, which coincides with a spike in total spin $\braket{\h{\mathbf{S}}^2}$. This was also highlighted in the Hamiltonian reconstruction metric applied to CNN and RBM NQS in Ref.~\cite{zhang2021hamiltonian}, which highlights the importance of respecting SU(2) symmetry in this region. Contrary to the latter type of models, we restrict our variational space to the manifold in Hilbert space of total SU(2) singlets and lattice-symmetry respecting wave functions, which clearly shows its benefit near $J_2/J_1=0.55$. This restriction is also imposed via the method of symmetry restoration from Ref.~\cite{Choo18}, and the analysis on RBMs in Ref.~\cite{nomura2020helping} (RBM+symm). The method described in Ref.~\cite{nomura2020helping} is able to separate the ground and lowest $S=1$ excited state by separating a symmetric and anti-symmetric part in the wave function with respect to spin flips. In our methodology, however, due to the inherent SU(2) conserving architecture, we are able to go beyond these two types of angular-momentum states, and can optimize for any total-angular momentum $S$. Compared to DMRG with built-in SU(2) symmetry, ClebschTree tends to obtain consistently lower ground-state energies.

One of the most interesting features of our model is that, compared to most other methods employing VMC (for example, Refs.~\cite{choo2019two, nomura2020helping}), we do not need to impose the Marshall sign rules, and are still able to obtain state-of-the-art approximations of the ground state. For example, our architecture obtains a far better approximation compared to Ref.~\cite{choo2019two}, where two types of sign rules are imposed. Hence, our method is able to efficiently capture these sign structures.

Furthermore, we point out that contrary to the Slater determinants for fermionic systems method (see Ref.~\cite{hu2013direct} with $p=0$), our method is systematically improvable via the hyperparameters $\Jmax$ and $\tau$, and therefore is able to reach a significantly lower ground state energy. Ref.~\cite{hu2013direct} addresses this issue by adding additional variational Lanczos steps. Ref.~\cite{ferrari2019neural} aims to overcome the same issue by modeling the correlation terms through a systematically improvable (but SU(2)-breaking) variational neural RBM ansatz. A similar approach is taken in Ref.~\cite{nomura2020dirac} (RBM+PP). Although this method appears to be very effective in capturing the ground state, it fails at capturing the challenging dynamics at $J_2/J_1 = 0.55$.

\subsection{Correlation functions}
\begin{figure}[tbh]
    \centering
    \includegraphics[scale=0.8]{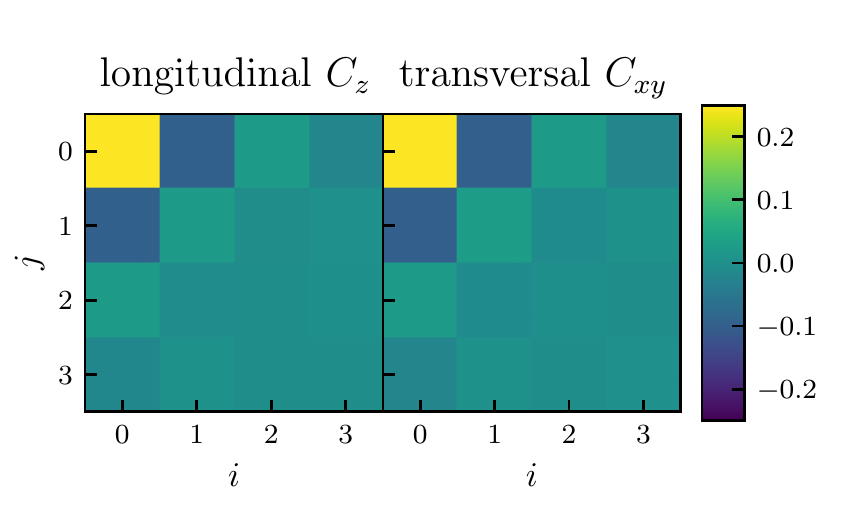}
    \caption{(left) Longitudinal and (right) transversal correlation functions as a function of the relative distance between spins for an $8\times8$ ground state of the \jj model at $J_2=0.5$.  Both components are equal, as is required by SU(2) invariance.}
    \label{fig:corrs}
\end{figure}
To highlight the symmetry in our variational ansatz, we now turn to correlation functions.  Our method conserves total angular momentum by construction.  For ground states of the models we investigate in this paper, this means that correlations between spin projections in the $z$-direction (longitudinal) and correlations between spin projections in the $xy$-plane (transversal) should be equal, as dictated by the total angular momentum.  The longitudinal and transversal correlation functions are respectively defined as 
\begin{align}
    C_z(i,j) &= \frac{1}{N} \sum_{k,l} \langle \h{\vec{s}}_{(k,l)}^z . \h{\vec{s}}_{(k+i,l+j)}^z \rangle,\label{eq:Cz}\\
    C_{xy}(i,j) &= \frac{1}{N} \sum_{k,l} \frac{1}{2} \langle \h{\vec{s}}_{(k,l)}^x . \h{\vec{s}}_{(k+i,l+j)}^x + \h{\vec{s}}_{(k,l)}^y . \h{\vec{s}}_{(k+i,l+j)}^y \rangle,\label{eq:Cxy}
\end{align}
where the sums over $k$ and $l$ run over all lattice positions and the correlations only depend on the relative position between the two spins due to periodic boundary conditions.

\begin{figure}[bth]
    \centering
    \includegraphics[scale=0.8]{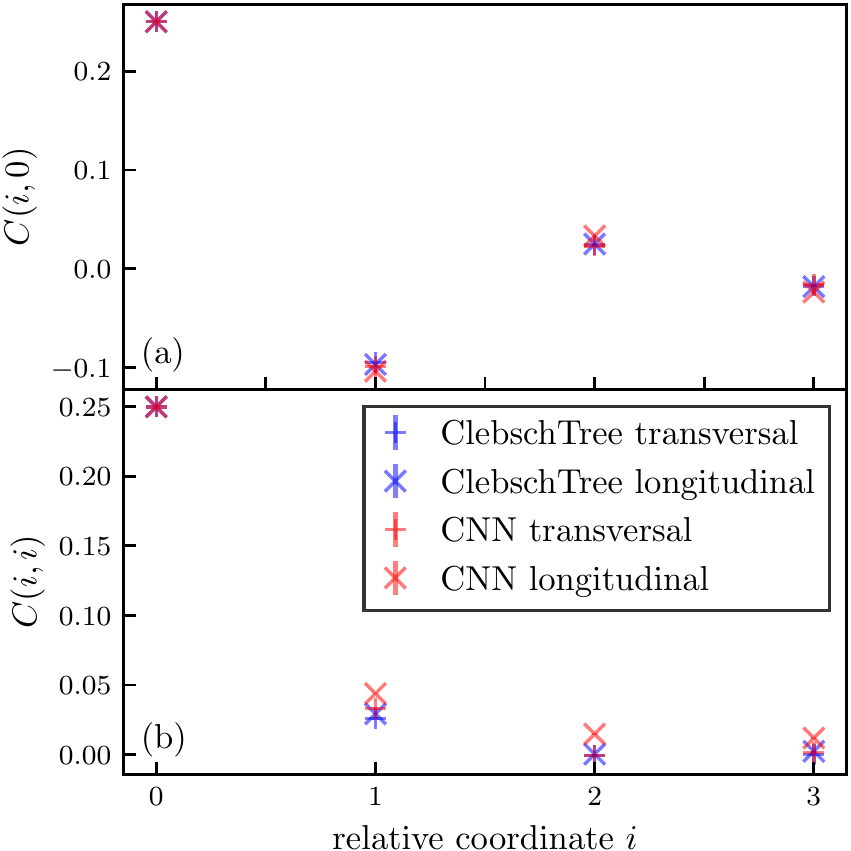}
    \caption{Correlation functions of our ansatz compared to those of a convolutional neural network (CNN) as in Ref.~\cite{choo2019two}. We plot the correlation functions between a fixed spin and those positioned horizontally (a) and diagonally (b) from it. The CNN breaks SU(2) invariance, while ClebschTree respects SU(2) symmetry by construction.}
    \label{fig:corrs_diff}
\end{figure}

To show that they are equal, we numerically compute both components of the spin-spin correlation function for our optimized ground state of the $8 \times 8$ \jj model, at $J_2 = 0.5$.  This is shown in Fig.~\ref{fig:corrs}, where we show both components of the correlation function for all possible relative positions of spins.  From this figure, it is indeed clear that the correlations behave as dictated by the angular momentum $S=0$.  In Fig.~\ref{fig:corrs_diff} we compare the transversal and longitudinal correlation functions of our ansatz versus those of a convolutional neural network.  Specifically, we look at the correlation between a fixed spin and other spins in two different directions, namely the horizontal direction and the diagonal direction.  The CNN breaks SU(2) as was shown in Ref.~\cite{choo2019two}.  This is also reflected in the difference between longitudinal and transversal correlation functions.  For our ansatz, both correlation functions are equal, up to small differences due to finite sample size in the estimation of the expectation values in Eqs.~\eqref{eq:Cz} and \eqref{eq:Cxy}.

\subsection{Excited states}
\begin{figure}[bth]
    \centering
    \includegraphics[width=0.5\textwidth]{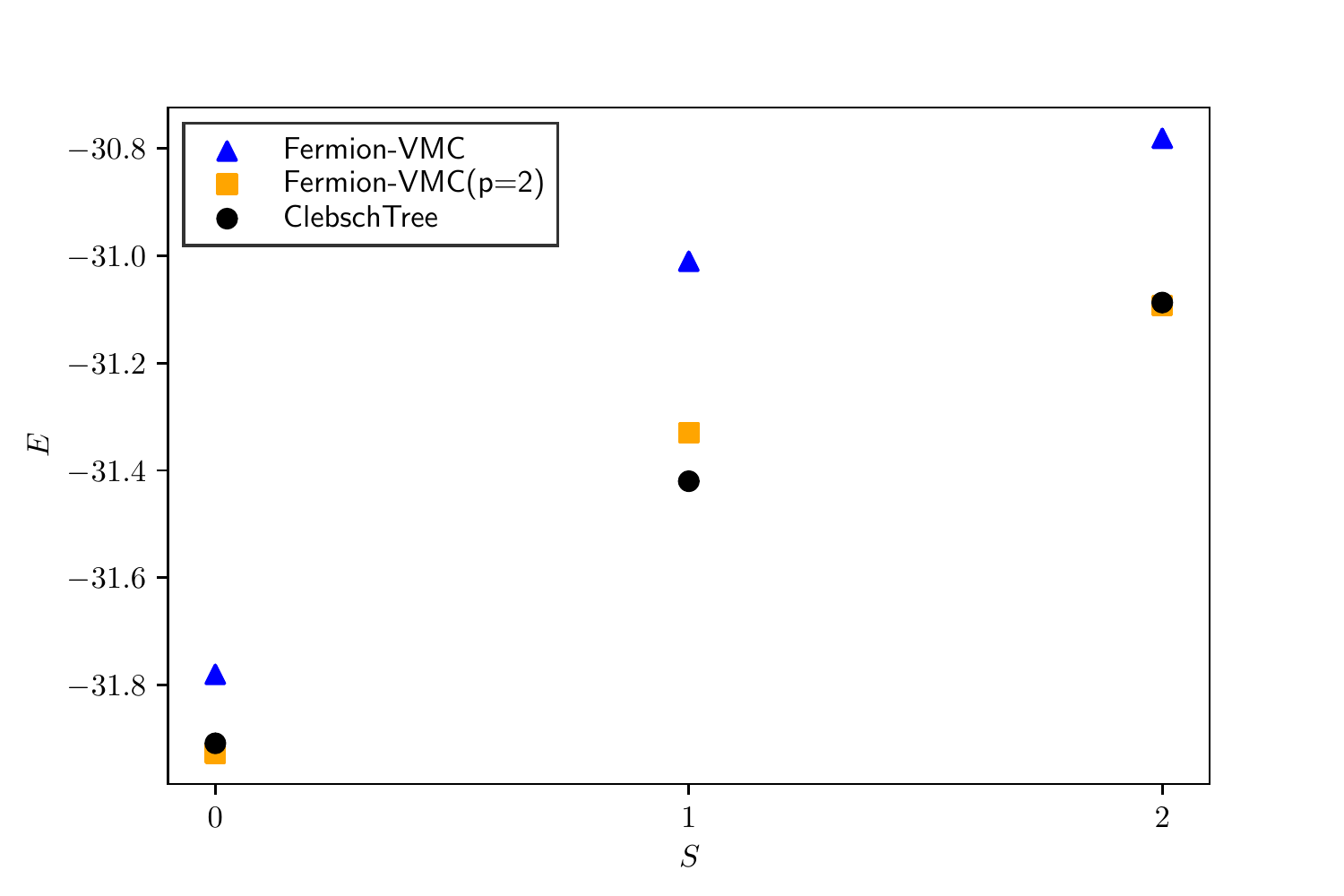}
    \caption{Energy of lowest states as a function of total angular-momentum $S$, for $J_2/J_1 = 0.5$.}
    \label{fig:excitations}
\end{figure}

As mentioned earlier, the symmetry restoration method from Refs.~\cite{choo2019two, nomura2020helping} can impose limited restrictions on the spin quantum number of the optimized state, because the singlet (triplet)
state is the lowest-energy state for each even (odd) total angular momentum $S$. In our method, the ansatz can be used to extract a wave function with any well-defined total angular momentum (including $S>1$). The results for $S=0,1,2$ are shown in Fig.~\ref{fig:excitations} and compared to Ref.~\cite{hu2013direct}. For the $S=1$ state, we isolate the $\mathbf{k} = (\pi,\pi)$ state in the $B_2$ irrep of the $C_{4v}$ lattice symmetry group, by choosing the correct characters in Eq.~\ref{eq:lattice_symmetries} (see detailed discussion in Ref.~\cite{nomura2020helping}). Notice that the latter approach, where we are able to select irreps of the lattice symmetry, results in far better energies at $S=1$.

\section{Conclusions and Outlook} In this paper, we presented a variational wave function and optimization scheme which is able to capture simultaneously continuous non-abelian symmetries and discrete lattice symmetries. By combining both types of symmetries we showed that excellent results can be obtained on the prototypical \jj model in two dimensions with periodic boundary conditions. The approach also allows to target different symmetry sectors than the ground state, and thus obtain excited states with definite quantum numbers with high accuracy. In our simulations, we did not observe the so-called sign-structure problem which is usually found in variational Monte Carlo studies, and therefore our work may provide an interesting example to study how this sign problem can be mitigated in future work.

Future directions using our approach include resolving the excitation spectrum of the \jj model at different values of the momentum and angular momentum quantum numbers in detail.  This would allow one to observe sudden changes in the structure of the spectrum and thus make claims on the nature of the \jj phase diagram. Other interesting avenues for further research are exploring the effect of coupling structures other than alternating nearest neighbour coupling. This can be especially useful for models where the ground state exhibits a valence bond structure with singlets spanning larger distances than neighbouring sites. On the architectural side, an interesting extension would be to find other efficient non-linear functions which allow the variational wave function to remain equivariant under SU(2), while providing us with more efficient parametrizations of quantum states. 
Although we considered square lattices and two dimensions in the current work, our model can also naturally be extended to different topologies and higher dimensions. For example, our method can form a powerful tool to study the \jj and Heisenberg model on e.g.\ triangular and face-centered qubic lattices to study geometrical frustration. Another application is fermionic models, where SU(2) symmetry is especially useful for e.g. Hubbard models or quantum chemistry problems.

\begin{acknowledgements}
We especially thank G.~Carleo for useful discussions and introducing Ref.~\cite{kondor2018clebsch} to us. We thank C.~Casert and J.~Ryckebusch for their useful comments on the final draft. We thank Y.~Nomura and K.~Choo for providing results from their models on the $8\times 8$ lattice. The software for this work was built on the \textsc{NetKet} library~\cite{carleo2019netket}. The computational resources (Stevin Supercomputer Infrastructure) and services used in this work were provided by the VSC (Flemish Supercomputer Center), and the Flemish Government -- department EWI. This work was supported by Ghent University, Special Research Fund Ghent University (J.~Nys), and Research Foundation Flanders (FWO-Flanders). The main body of this work was carried out while J.~Nys was at Ghent University. T.~Vieijra is supported as an `FWO-aspirant' under contract number FWO18/ASP/279.
\end{acknowledgements}

\bibliography{references.bib}

\revappendix

\section{Numerical results}\label{app:numval}

In Table~\ref{tab:results8x8}, we provide the numerical results related to \linktomain{
Fig.~2
in the main text}.

\begin{table*}[tbh]
    \centering
    \caption{Results for 8x8 lattice for various $J_2/J_1$.}
    \label{tab:results8x8}
    \begin{tabular}{c|c|c|c|c|c}
    	\hline\hline 
	& \multicolumn{5}{c}{$J_2/J_1$}\\ 
	\hline
Reference	& $0$	& $0.45$	& $0.5$	& $0.55$	& $1$\\ 
	\hline 
CNN \cite{choo2019two}	& $-43.09275 \pm 0.00050$	& $-32.64850 \pm 0.00100$	& $-31.76950 \pm 0.00125$	& $-30.98850 \pm 0.00150$	& -\\ 
DMRG-SU(2) \cite{gong2014plaquette}	& -	& $-32.68736 \pm 0.00000$	& $-31.88320 \pm 0.00000$	& $-31.24224 \pm 0.00000$	& -\\ 
Fermion-VMC \cite{hu2013direct}	& -	& $-32.59520 \pm 0.00064$	& $-31.77984 \pm 0.00064$	& $-31.03168 \pm 0.00064$	& -\\ 
Fermion-VMC(p=2) \cite{hu2013direct}	& -	& $-32.72000 \pm 0.00064$	& $-31.92704 \pm 0.00064$	& $-31.25824 \pm 0.00128$	& -\\ 
RBM+PP \cite{nomura2020dirac}	& -	& $-32.71149 \pm 0.00026$	& $-31.90144 \pm 0.00038$	& $-31.21984 \pm 0.00064$	& -\\ 
RBM+symm \cite{nomura2020helping}	& $-43.10268 \pm 0.00003$	& -	& $-31.91460 \pm 0.00015$	& -	& -\\ 
ClebschTree (Ours)	& $-43.09640 \pm 0.00320$	& $-32.70150 \pm 0.00590$	& $-31.90850 \pm 0.00420$	& $-31.25440 \pm 0.00510$	& $-44.75400 \pm 0.01800$\\
	\hline\hline 
    \end{tabular}
\end{table*}

\section{Optimization details}\label{app:optimization}
Observables are estimated by evaluating the expectation values through Markov Chain Monte Carlo sampling. For example, using a set of Monte Carlo configurations $\mathcal{S}=(m_{s_1}, ... , m_{s_N})$ in a spin-projection basis, generated via the distribution $\left|\Psi(\mathcal{S})\right|^2$, the energy can be estimated via~\cite{carleo2019netket}
\begin{align}
    \braket{\hat{H}} = \left< \sum_{\mathcal{S'}} \bra{\mathcal{S}} \hat{H} \ket{\mathcal{S'}} \frac{\Psi(\mathcal{S'})}{\Psi(\mathcal{S})} \right>_{\left|\Psi(\mathcal{S})\right|^2}\label{eq:energy_expectation}
\end{align}
We use $4096$ samples to estimate the expectation values in each step. Automatic differentiation~\cite{jax2018github} via the Jax library backpropagates the gradients through our network, and we use the Adam optimizer with learning rate $0.001$. Furthermore, we use gradient clipping with a maximum gradient norm of $0.001$. To find the global optimum for the parameters, we initially use a different procedure to quickly determine the region of parameter space where the ground state is located. Therefore, we first carry out a Stochastic Reconfiguration (SR) step with a uniform diagonal shift of magnitude $0.01$, followed by a step using the resulting update direction with an Adam optimizer. To save computational resources, in this optimization regime we use $256$ samples and learning rate $0.005$. Afterwards, we switch to the standard optimization strategy (and drop the SR step). For the Adam optimizer, we use hyperparameters $\beta_1=0.9$ and $\beta_2=0.999$. Especially for the excitation energies, the energy obtained in this first step is close to the final result. The number of connected configurations $\mathcal{S'}$ in Eq.~\eqref{eq:energy_expectation} can become quite large in the case of many samples $\mathcal{S}$, and therefore, to restrict the memory usage, the forward and backward passes are batched. 
 
We pay specific attention to the initialization of the variational parameters in our model.  As the nature of our operations is a sum of multiplications, our model is prone to numerical under- and overflow, especially for larger systems.  Furthermore, the issue of vanishing/exploding gradients arises when the covariances of intermediate features grow or diminish with increasing depth. First, we initialize the weight matrices $W$ via a uniform distribution. Next, we compute the activations in a forward pass through a layer for a set of $1000$ random configurations $\mathcal{S}$. We then rescale the original random weight matrices $W$ by the computed standard deviations of the activations of that layer to obtain a new matrix $W'$. Next, the activations are recomputed using $W'$, and passed to the next layer to perform a similar rescaling of the weights in that layer. In such a way, we obtain an activation variance close to $1$ in each layer during the first step of the optimization problem, thereby greatly reducing the chance of under- or overflow.
\end{document}